\documentclass[conference]{IEEEtran}
\IEEEoverridecommandlockouts

\usepackage{cite}
\usepackage{amsmath,amssymb,amsfonts}
\usepackage{graphicx}
\usepackage{textcomp}
\usepackage{xcolor}
\usepackage{acro}

\DeclareAcronym{SLA}{
    short = SLA ,
    long  = Service Level Agreement
}

\DeclareAcronym{SixG}{
    short = 6G ,
    long  = Sixth Generation
}

\DeclareAcronym{XR}{
    short = XR ,
    long  = Extended Reality
}

\DeclareAcronym{KPI}{
    short = KPI ,
    long  = Key Performance Indicator
}

\DeclareAcronym{KVI}{
    short = KVI ,
    long  = Key Value Indicator
}

\DeclareAcronym{KV}{
    short = KV ,
    long  = Key Value
}

\DeclareAcronym{ICT}{
    short = ICT ,
    long  = Information and Communication Technology
}

\DeclareAcronym{RD}{
    short = R\&D ,
    long  = Research and Development
}

\DeclareAcronym{VoS}{
    short = VoS ,
    long  = Value of Service
}

\DeclareAcronym{BI}{
    short = BI ,
    long  = Business Intelligence
}

\DeclareAcronym{TRL}{
    short = TRL ,
    long  = Technology Readiness Level
}

\DeclareAcronym{TOPSIS}{
    short = TOPSIS ,
    long  = Technique for Order Preference by Similarity to Ideal Solution
}

\DeclareAcronym{SDG}{
    short = SDG ,
    long  = Sustainable Development Goal
}

\DeclareAcronym{ISO}{
    short = ISO ,
    long  = International Organization for Standardization
}

\DeclareAcronym{UI}{
    short = UI ,
    long  = User Interface
}

\DeclareAcronym{API}{
    short = API ,
    long  = Application Programming Interface
}

\DeclareAcronym{LLM}{
    short = LLM ,
    long  = Large Language Model
}

\DeclareAcronym{SLM}{
    short = SLM ,
    long  = Small Language Model
}

\DeclareAcronym{CoT}{
    short = CoT ,
    long  = Chain-of-Thought
}

\DeclareAcronym{RAG}{
    short = RAG ,
    long  = Retrieval-Augmented Generation
}

\usepackage{fancyhdr,lipsum}

\fancypagestyle{mahmood}{%
   \fancyhf{} 
   
   \fancyhead[C]{\textit{You can use this material personally. Reprinting or republishing this material for the purpose of advertising or promotion, creating new collective works, reselling or redistributing to servers or lists, or using any copyrighted component in other works must adhere to IEEE policy.}}
}%

\makeatletter
\let\ps@IEEEtitlepagestyle\ps@mahmood
\makeatother

\begin{document}

\title{KPI2KVI: A Multi Agent Workflow for Calculating Key Value Indicators from Service Descriptions}

\author{
    \IEEEauthorblockN{Masoud Shokrnezhad}
    \IEEEauthorblockA{\textit{ICTFICIAL OY}\\
        Espoo, Finland\\
        masoud.shokrnezhad@ictficial.com}
    \and
    \IEEEauthorblockN{Tarik Taleb}
    \IEEEauthorblockA{\textit{Ruhr-Universitaet Bochum}\\
        Bochum, Germany\\
        tarik.taleb@ruhr-uni-bochum.de}
    \and
    \IEEEauthorblockN{Yan Chen}
    \IEEEauthorblockA{\textit{ICTFICIAL OY}\\
        Espoo, Finland\\
        yan.chen@ictficial.com}
    \and
    \IEEEauthorblockN{Qize Guo}
    \IEEEauthorblockA{\textit{ICTFICIAL OY}\\
        Espoo, Finland\\
        qize.guo@ictficial.com}
}

\maketitle

\begin{abstract}
    Key Value Indicators (KVIs) provide a decision oriented view of a service by summarizing how operational performance translates into stakeholder value, risk, and outcomes. However, in many domains KVIs are difficult to compute in practice because they require selecting relevant KVI categories, defining measurable Key Performance Indicators (KPIs), collecting KPI values, and applying consistent calculation logic, all of which is typically performed manually and inconsistently from unstructured service documentation. This paper presents KPI2KVI, a tool that transforms a natural language service description into computed KVI estimates by orchestrating a deterministic multi agent workflow powered by Large Language Models (LLMs) that (i) elicits missing service context, (ii) extracts and finalizes relevant KVI categories from a taxonomy, (iii) generates service specific KPIs with units and descriptions, (iv) collects KPI values through an interactive dialogue and also supports intelligent estimation for KPI values that are unavailable, and (v) computes interval valued KVI outputs (minimum, exact, maximum) with traceable explanations for each KVI code. Simulations with representative service descriptions demonstrate that KPI2KVI consistently produces a complete end to end mapping from description to KVI intervals and provides transparent calculation narratives that support post hoc auditing and interactive advisory queries.
\end{abstract}


\section{Introduction}
The \ac{SixG} vision increasingly frames networks as critical societal infrastructure, expected not only to deliver advanced capabilities but also to contribute to long-term objectives such as sustainability, inclusion, resilience, and trust \cite{wikstrom_what_2022, wikstrom_key_2024, wang_6gpath_2026, farhoudi_service_2025}. As these expectations shape research agendas, governance, and procurement, stakeholders need ways to \emph{demonstrate}, \emph{compare}, and \emph{audit} the value impact of services (not just their technical performance) across heterogeneous contexts and lifecycle stages \cite{pintor_sustainability_2025, farhoudi_service_2025}. \acp{KVI} were proposed to make such value outcomes measurable and actionable, but in practice their computation is challenging: value effects are indirect and multi-stakeholder, relevant evidence is often incomplete or only available via proxies (measurements, certifications, surveys), and early-stage designs must still provide credible estimates with transparent assumptions \cite{wikstrom_what_2022}. Without reproducible calculation workflows that link service descriptions to evidence-backed \acp{KVI}, value assessment risks becoming ad-hoc, hard to optimize, and prone to inconsistency or ``value-washing'' \cite{wikstrom_key_2024}.

Existing approaches broadly fall into two strands. First, concept and governance frameworks motivate \acp{KVI} and propose value-driven assessment processes (e.g., eliciting stakeholders and values, defining indicator candidates, and staging evaluation by maturity), which is valuable for shared language and decision-making, but typically stops short of prescribing fully specified, end-to-end computation pipelines from service descriptions to concrete, reproducible indicator values. Second, operational approaches embed \acp{KVI} into orchestration and optimization (e.g., ranking alternatives or trading off performance and value objectives), demonstrating that value-aware decisions are possible when indicators are computable, but often do so for a small, pre-selected \ac{KVI} set and under strong assumptions about the availability, meaning, and provenance of required inputs. In practice, however, the hard part is often \emph{operationalization} under realistic constraints: deciding which \acp{KVI} are actually relevant for a new heterogeneous service, turning narrative requirements into a measurement plan, and then computing results from mixed evidence sources where some inputs are missing, approximate, or only available via proxies. These issues motivate end-to-end traceability and uncertainty-aware outputs with clear, user-facing rationales.

To address these gaps, this paper proposes KPI2KVI, a \ac{LLM}-powered multi-agent workflow that computes \acp{KVI} from a service description in a general, reproducible, and traceable way. KPI2KVI uses specialized \ac{LLM}-based agents to (i) conduct a guided interview that elicits service intent, context, stakeholders, and potential value impacts, (ii) map the service to a controlled \ac{KVI} taxonomy and finalize the \ac{KVI} scope with human-in-the-loop refinement, (iii) generate a compact, service-specific \ac{KPI} evidence plan for the selected \acp{KVI} and collect/structure the resulting measurements with provenance, and (iv) compute each \ac{KVI} with explicit \(\{\text{exact},\text{min},\text{max}\}\) bounds and a short rationale that cites the precise \acp{KPI} inputs and assumptions used. By combining \ac{LLM} semantic understanding with a deterministic staged pipeline and persistent structured artifacts, KPI2KVI systematically bridges stakeholder value expectations to measurable evidence and makes uncertainty, assumptions, and computation steps explicit and auditable.

The rest of this paper is organized as follows. Section~II reviews some related work on \acp{KVI} concepts, frameworks, and optimization-based operationalizations. Section~III presents the KPI2KVI workflow and architecture in detail. Section~IV evaluates the approach through simulations. Section~V concludes and outlines directions for future work.

\section{Literature Review}
In \ac{SixG}, a \emph{service} is an end-to-end capability for users or verticals realized by chaining functions across heterogeneous domains (edge/cloud, terrestrial/non-terrestrial) and governed through \acp{SLA} or intent-based abstractions \cite{de_trizio_optimizing_2024,wikstrom_what_2022,farhoudi_deep_2025,farhoudi_service_2025}. A service request therefore combines functional goals with workload characteristics and stringent \emph{requirements} on latency, throughput/data rate, reliability/availability, coverage, positioning, privacy, and security, often context-dependent \cite{de_trizio_optimizing_2024,wikstrom_what_2022,shokrnezhad_near-optimal_2022,shokrnezhad_joint_2018,mazandarani_novel_2025,mazandarani_semantic-aware_2025,mazandarani_adaptive_2025}. In intent-based formulations, services may be decomposed into tasks and mapped to intent categories to make requirements machine-actionable and comparable \cite{de_trizio_optimizing_2024}. \acp{KPI} are technical, measurable quantities estimating performance (e.g., delay, throughput, packet loss), typically specified as desired values with tolerable thresholds \cite{sciddurlo_value-driven_2025,wikstrom_what_2022}. \acp{KVI} complement \acp{KPI} by estimating enabled (or harmed) societal values such as sustainability, inclusion, privacy/confidentiality, and trust \cite{wikstrom_what_2022,wikstrom_key_2024}. Because many value dimensions are not directly observable at run time, \acp{KVI} are often realized via sensor-based measurements, periodic certification/audits, and compositions over lower-level indicators \cite{pintor_sustainability_2025}. This \ac{KPI}--\ac{KVI} split motivates values-driven service design and evaluation beyond performance-only engineering.

Several works defined \acp{KVI} and provided high-level workflows rather than fully specified calculation pipelines. Atzori \textit{et al.} \cite{atzori_toward_2023} proposed EthicNet/\ac{VoS}, where stakeholders expressed \ac{KVI} requirement profiles and \acp{KVI} should be monitored and composed end-to-end, but composition operators remained open. Pintor \textit{et al.} \cite{pintor_sustainability_2025} (and \cite{pintor_building_2024}) systematized architectural formalization (sensor-based vs.\ certified \acp{KVI}, metadata, and parent--child structuring). Wikstr\"om \textit{et al.} \cite{wikstrom_key_2024} offered a five-step framework from scenario/value elicitation to \ac{KVI} formulation and staged assessment, while their white paper defined \acp{KVI} as the \emph{scale of effect} of a use case and linked \acp{KV}$\rightarrow$\acp{KVI}$\rightarrow$enablers$\rightarrow$\acp{KPI}, explicitly noting the unknown ``exchange rate'' between usage and societal value \cite{wikstrom_what_2022}. Ziegler \textit{et al.} \cite{ziegler_6g_2020} provided qualitative \ac{KPI}-to-value impact mapping, and Osman \textit{et al.} \cite{osman_bridging_2024} emphasized \ac{KPI} proxies and business-model-driven prioritization. Other \ac{SixG} discussions highlighted ecosystem/governance drivers without prescribing calculation rules \cite{seppo_yrjola_value_2022,christophorou_adroit6g_2023,pouttu_6g_2020}, while enterprise \ac{BI} work focused on \ac{KPI} aggregation/visualization rather than \ac{KPI}$\rightarrow$\ac{KVI} translation \cite{kolychev_application_2019}. Across these contributions, assessment was often staged by maturity: early \acp{TRL} relied on expert/qualitative evidence, later \acp{TRL} relied on measurement and more objective signals \cite{wikstrom_what_2022}.

A second group operationalized \acp{KVI} through explicit calculation embedded in orchestration/optimization. De Trizio \textit{et al.} \cite{de_trizio_optimizing_2024} modeled intent-mapped service provisioning as a many-to-many matching problem, combining \ac{KPI} constraints (deadline, throughput) with \ac{KVI}-related constraints (budget, risk appetite) and ranking providers by entropy-weighted \ac{TOPSIS} over cost and cyber risk. Sciddurlo \textit{et al.} \cite{sciddurlo_value-driven_2025} defined \ac{KPI} vectors for services/resources, computed \ac{KVI} components (environmental sustainability, trustworthiness, inclusiveness) via formulas, and optimized \ac{KPI}--\ac{KVI} trade-offs via a bi-objective model solved by an exact $\epsilon$-constraint method; they also proposed translating natural-language requests into intents enriched with \acp{KPI} and \acp{KVI}. Mertens \textit{et al.} \cite{mertens_deep_2024} proposed \ac{SDG}-indexed \acp{KVI} where ``objective'' service \acp{KVI} could be scored via \ac{ISO}-standards coverage ratios and combined with user preference profiles.
Methodologically, these works relied on normalization and aggregation (e.g., relative-closeness ranking in \cite{de_trizio_optimizing_2024}, weighted-sum \ac{KVI} aggregation and Pareto optimization in \cite{sciddurlo_value-driven_2025}).

Despite rapid progress, gaps remained for computing \emph{service-relevant} \acp{KVI} in a general, reproducible way. Vision/framework works clarified \ac{KVI} concepts and governance, but often did not specify how to select relevant \acp{KVI} for a new service nor how to compute them from available evidence without extensive manual modeling \cite{atzori_toward_2023,wikstrom_key_2024,pintor_sustainability_2025}. Qualitative mappings and vertical matrices communicated priorities but were difficult to audit, compare, or optimize because scales and aggregation rules were coarse or subjective \cite{ziegler_6g_2020,pouttu_6g_2020}. Optimization-oriented work, in contrast, typically fixed a small \ac{KVI} set and assumed access to non-trivial inputs (e.g., carbon factors, attack likelihoods, certification mappings) that might be unavailable or ambiguous at design time \cite{de_trizio_optimizing_2024,sciddurlo_value-driven_2025,mertens_deep_2024}. Across strands, there was limited support for missing/uncertain evidence, end-to-end traceability from service description to computed indicators, and systematic bridging from stakeholder value expectations to measurable \ac{KPI} evidence \cite{wikstrom_what_2022,osman_bridging_2024}. Moreover, many approaches provided neither uncertainty bounds (e.g., intervals) nor concise, user-facing rationales connecting results to underlying evidence. These limitations motivated more operational and traceable \ac{KPI}-to-\ac{KVI} computation workflows that remained usable under uncertainty and heterogeneous services.

\section{Approach}
This section presents KPI2KVI, a multi-agent workflow that calculates \acp{KVI} for a given service. We first define the nine-stage workflow and the roles of its \ac{LLM}-based agents, then describe the system architecture and orchestration logic that implement this workflow, and finally walk through a concrete example that shows how a small set of \acp{KVI} can be calculated for a cloud-based telemedicine video consultation service.

\begin{figure*}[t]
    \centering
    \includegraphics[width=\textwidth]{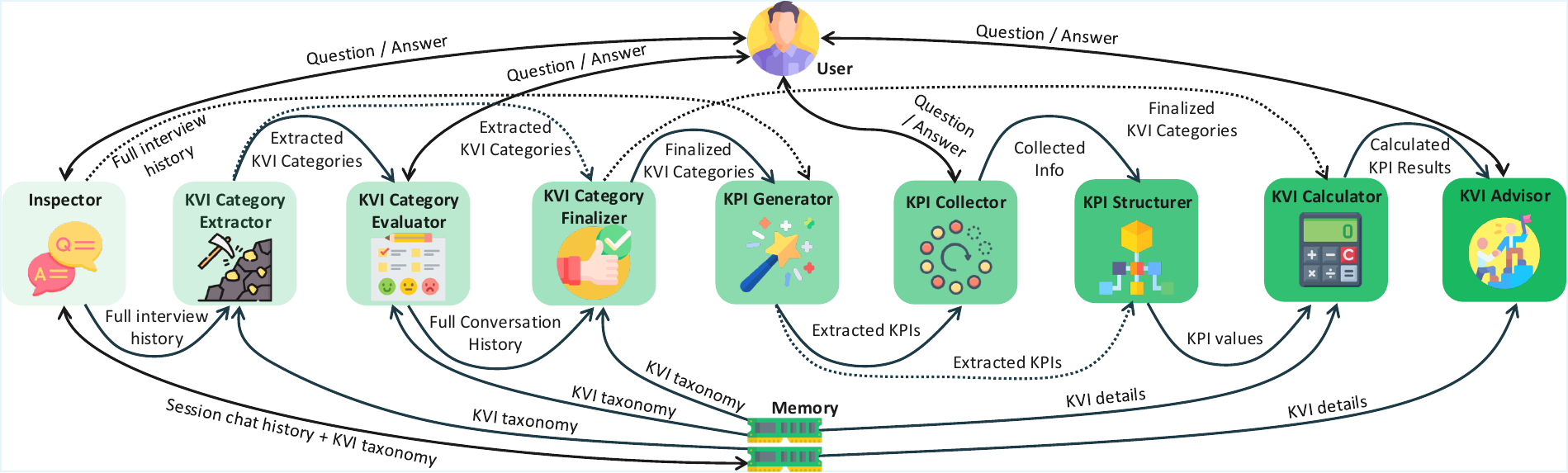}
    \caption{KPI2KVI pipeline: from service interview to \ac{KVI} category selection, \ac{KPI} generation and value collection, per-\ac{KVI} calculation with explicit bounds, and post-hoc advising, with shared memory storing reusable artifacts across stages.}
    \vspace{-0.5cm}
    \label{fig:kpi2kvi-workflow}
\end{figure*}

\subsection{Workflow and Agent Responsibilities}
The end-to-end workflow is a deterministic pipeline composed of nine steps, alternating between conversational steps (to elicit information and allow corrections) and structured steps (to produce machine-readable artifacts). The progression of stages and the shared memory that carries artifacts across them are shown in Fig.~\ref{fig:kpi2kvi-workflow}.

\paragraph{\ac{KVI} Category Selection} The workflow starts with the \emph{inspector} (Step~1), whose \textit{inputs} are the initial service description and any subsequent clarifications. Its \textit{process} is a guided interview that elicits service intent, operational context, stakeholders, and potential value impacts (e.g., privacy/security expectations, sustainability concerns, accessibility constraints), and its \textit{output} is an interview transcript stored in memory and used as the primary evidence for downstream selection. Next, the \emph{kvi category extractor} (Step~2) takes as \textit{inputs} the full inspector transcript and the global \ac{KVI} taxonomy, \textit{processes} them by mapping the service to the most relevant \ac{KVI} categories using the taxonomy as a controlled vocabulary, and \textit{outputs} a ranked structured list of candidate categories for refinement. The taxonomy is extracted from the KVIs proposed in \cite{patsouras_6g_2025}. The \emph{kvi category evaluator} (Step~3) then uses as \textit{inputs} the extracted categories and taxonomy context, \textit{processes} them via a short conversational loop that justifies the proposal and invites the user to add/remove/replace categories based on domain knowledge and stakeholder priorities, and \textit{outputs} a refinement transcript capturing the user’s decisions. Finally, the \emph{kvi category finalizer} (Step~4) takes as \textit{inputs} the extractor output and evaluator transcript, \textit{processes} them by consolidating the final category set (resolving duplicates and ensuring consistent IDs), and \textit{outputs} a structured finalized list of categories that defines the \ac{KVI} scope for subsequent \ac{KPI} generation and \ac{KVI} calculation; this contract is critical for auditability because it fixes \emph{which} values will be computed before numerical evidence is collected.

\paragraph{KPI Generation} Given the finalized \ac{KVI} scope, KPI2KVI next produces the measurement evidence needed for computation. The \emph{kpi generator} (Step~5) takes as \textit{inputs} the inspector transcript (service context) and the finalized \ac{KVI} categories (value scope), \textit{processes} them by proposing a compact set of service-specific \acp{KPI} that can serve as measurable evidence for the selected value indicators (including name, description, and unit for each \ac{KPI}), and \textit{outputs} a structured \ac{KPI} list stored in memory and presented to the user as the measurement plan. The \emph{kpi collector} (Step~6) then uses as \textit{inputs} this \ac{KPI} list and the ongoing user chat, \textit{processes} them by collecting one \ac{KPI} value at a time while explicitly supporting missing evidence (the user may provide a value or delegate it to the system, which is recorded as an assumption rather than an observation), and \textit{outputs} a collection transcript capturing raw values, units, and whether each value was user-provided or system-decided. Finally, the \emph{kpi structurer} (Step~7) takes as \textit{inputs} the \ac{KPI} list and the collector transcript, \textit{processes} them by converting free-form conversation into a machine-readable table of \ac{KPI} values with provenance flags and normalized representation (e.g., consistent numeric parsing and unit alignment), and \textit{outputs} a structured \ac{KPI} table that becomes the single source of truth for downstream \ac{KVI} calculations.

\paragraph{KVI Calculation} Given the structured \ac{KPI} table and finalized \ac{KVI} scope, KPI2KVI computes the target indicators and makes the results explainable. The \emph{kvi calculator} (Step~8) takes as \textit{inputs} (i) one target \ac{KVI} definition, including a step-by-step calculation narrative (a reasoning chain) describing how to derive the \ac{KVI} and (ii) the structured \ac{KPI} table, \textit{processes} them by producing an estimated value with explicit bounds and a short rationale that links the result to the exact \acp{KPI} and assumptions used, and \textit{outputs} per-\ac{KVI} calculation artifacts \(\{\text{exact},\text{min},\text{max},\text{rationale}\}\) stored in memory. The orchestrator loops this step over all \acp{KVI} implied by the finalized categories and stores results so later stages can cite them precisely. Finally, the \emph{kvi advisor} (Step~9) takes as \textit{inputs} a consolidated advisor context containing the finalized categories, the structured \ac{KPI} table, and all per-\ac{KVI} calculation artifacts, \textit{processes} them by answering user questions and explaining trade-offs with traceable justifications, and \textit{outputs} user-facing explanations and follow-up guidance (e.g., which \acp{KPI} would most reduce uncertainty if measured more precisely).

\subsection{Architecture}
\begin{figure}[t]
    \centering
    \includegraphics[width=\columnwidth]{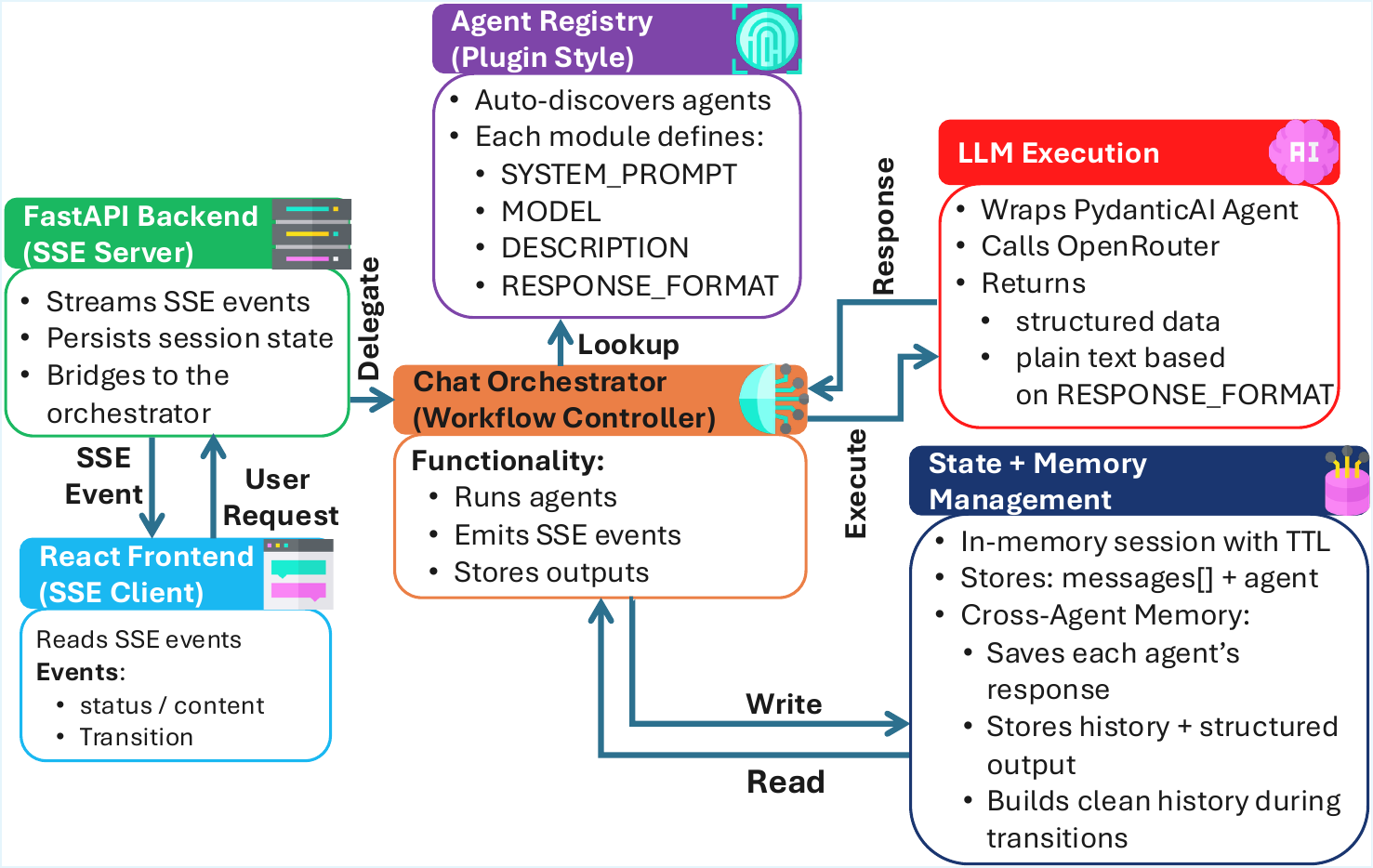}
    \caption{TheKPI2KVI architecture: a streaming frontend/back\-end setup with a workflow controller that discovers agents, executes them via an LLM provider, and manages cross-agent state and reusable structured artifacts.}
    \vspace{-0.5cm}
    \label{fig:kpi2kvi-arch}
\end{figure}

The KPI2KVI workflow is implemented as a two-tier system with a chat interface and a backend workflow controller. The frontend provides a conversational \ac{UI} and consumes a Server-Sent Events stream to incrementally render intermediate status updates and agent outputs. The backend exposes a streaming \ac{API} endpoint and delegates each user turn to a workflow engine that selects and executes the appropriate step of the workflow, persists the session state, and emits progress and content events for the UI. The runtime architecture is shown in Fig.~\ref{fig:kpi2kvi-arch}. The workflow controller (orchestrator) is the central component that realizes the nine-stage process described above. It discovers agent modules from a registry, executes them through an \ac{LLM} interface, and maintains cross-agent memory. Concretely, the orchestrator implements a staged state machine where each session stores (i) the complete chat history, (ii) the name of the \emph{current} agent responsible for the next user message, and (iii) a key--value store of artifacts accumulated so far. Artifacts include both human-readable text (assistant messages shown to the user) and structured objects (e.g., lists of selected \ac{KVI} category IDs, a \ac{KPI} list with units/descriptions, and a table of collected \ac{KPI} values with provenance). This design is what enables the system to advance from one workflow step to the next without losing context, while still keeping each step’s output explicit and auditable.

Operationally, the orchestrator performs three actions on every user turn. First, it \emph{builds context}: it constructs the next prompt by combining the new user message with the relevant subset of stored artifacts (e.g., inspector transcript for category extraction; finalized categories for \ac{KPI} generation; structured \ac{KPI} table for \ac{KVI} calculation). Second, it \emph{runs and routes}: it executes the current agent, inspects the response for completion cues (for conversational steps), and when a step is complete it automatically triggers the next structured step(s) without requiring another user message. Third, it \emph{stores and streams}: it writes outputs back to memory under explicit keys and streams progress/content events so the frontend can render multi-agent turns coherently. Finally, the architecture separates \emph{domain knowledge} from \emph{interaction logic}: the orchestrator loads the shared \ac{KVI} taxonomy and canonical \ac{KVI} definitions from data assets and passes them into relevant agents, ensuring stable IDs/codes and enabling reuse of intermediate structured artifacts across runs.

\subsection{Illustrative Example and Addressing Literature Gaps}
Consider a simple service: a cloud-based video consultation platform that enables remote medical appointments. The service description typically includes performance intent (stable video/audio, low delay, high availability) and operational constraints (sensitive personal data, multi-tenant deployment, and varying client devices) \cite{farhoudi_deep_2025,shokrnezhad_near-optimal_2022,mazandarani_novel_2025,mazandarani_adaptive_2025}. In KPI2KVI, the inspector first elicits missing context (e.g., what data are processed and stored, who can access recordings, which regulations apply, and the expected user base). Based on this transcript, the category extractor selects a single relevant \ac{KVI} category, for example \emph{User Trust, Perception, \& Requirement Compliance}, which in the KPI2KVI taxonomy maps to \acp{KVI} such as \texttt{PUC-UPCA} (share of identified user privacy concerns addressed, \%), \texttt{PUC-USCA} (share of identified user security concerns addressed, \%), and \texttt{RPS-DDSS} (user-reported perceived security of the service, \%).

The \ac{KPI} generator then proposes a small evidence set sufficient to compute these \acp{KVI}. For this example, consider three \acp{KPI}: \(N_p\) = number of privacy concerns collected during requirements elicitation (count), \(A_p\) = number of those privacy concerns addressed by implemented controls (count), and \(r_s\) = average user perceived-security score on a 1--5 Likert scale from a short pilot survey (dimensionless). Suppose the \ac{KPI} collection yields \(N_p = 10\), and because some controls are still under implementation the collector records \(A_p \in [7,9]\) (delegated estimate). From a pilot with limited respondents, assume \(r_s \in [3.8,4.4]\) with a nominal mean \(r_s = 4.1\). These values are stored in the structured \ac{KPI} table with provenance (the interval-valued entries explicitly reflecting uncertainty).

The \ac{KVI} calculator then produces transparent formulas and interval results. For \texttt{PUC-UPCA}, a natural operationalization is the percentage of addressed privacy concerns:
\[
    \texttt{PUC-UPCA} = 100 \cdot \frac{A_p}{N_p}.
\]
With \(N_p=10\) and \(A_p \in [7,9]\), the result is \(\texttt{PUC-UPCA}_{\min}=70\), \(\texttt{PUC-UPCA}_{\max}=90\), and \(\texttt{PUC-UPCA}_{\text{exact}}=80\) (using the midpoint). For perceived security \texttt{RPS-DDSS}, the calculator can map the Likert score to a 0--100 scale:
\[
    \texttt{RPS-DDSS} = 100 \cdot \frac{r_s-1}{4}.
\]
With \(r_s \in [3.8,4.4]\), we obtain \(\texttt{RPS-DDSS}_{\min}=70\), \(\texttt{RPS-DDSS}_{\max}=85\), and \(\texttt{RPS-DDSS}_{\text{exact}}=77.5\). Importantly, the advisor can later explain that the uncertainty comes from delegated control-coverage assumptions and limited survey evidence, and can recommend which \acp{KPI} (e.g., tightening \(A_p\) via control verification or increasing the survey sample for \(r_s\)) would most reduce the \ac{KVI} bounds. Overall, the example demonstrates a concrete chain from service description to bounded, explainable \acp{KVI} outputs grounded in an explicit \ac{KPI} evidence set.

\subsection{Addressing Literature Gaps}
KPI2KVI directly addresses the main gaps identified in Sec.~II by making \ac{KPI}-to-\ac{KVI} computation operational, traceable, and usable under uncertainty, while keeping the workflow largely automatic through \acp{LLM}. First, instead of assuming a fixed, small \ac{KVI} set or requiring extensive manual modeling, \acp{LLM} perform taxonomy-grounded category extraction and refinement, and the inspector/evaluator loops explicitly incorporate human feedback before a finalized category contract is recorded. Second, it systematically bridges stakeholder value expectations to measurable evidence by using \acp{LLM} to propose a compact, service-specific \ac{KPI} set and by structuring collected values into a single machine-readable table that can be reused and compared across runs. Third, computation and explanation are supported by the \ac{KVI} definitions and the calculator’s step-by-step reasoning: for each \ac{KVI}, the system produces bounded results with short rationales tied to the exact \acp{KPI} and assumptions, and the advisor can answer questions by citing stored artifacts. Together, these mechanisms improve end-to-end traceability from the service description to computed \acp{KVI}, and make uncertainty and human-in-the-loop corrections explicit rather than implicit.

\section{Simulations}
We evaluate KPI2KVI by testing four variants: (1) \emph{monolithic LLM} (DeepSeek-R1), where a single system prompt describes the end-to-end KPI2KVI logic and embeds the \ac{KVI} taxonomy; (2) \emph{agentic SLM (no taxonomy, no CoT)}, where the KPI2KVI agents are implemented with an \ac{SLM} (Gemini 2.5 Flash Lite) and have no access to the taxonomy and no explicit \ac{CoT} calculation prompting; (3) \emph{agentic SLM (+ taxonomy, no CoT)}, identical to (2) but with access to the taxonomy artifacts; and (4) \emph{KPI2KVI (full)}. To produce the measurements, we execute each method end-to-end on a suite of service cases and systematically vary three experimental conditions: (i) the computational difficulty of \ac{KVI} derivation, (ii) the scope of requested \acp{KVI}, and (iii) the quality of taxonomy grounding available to the method. Concretely, calculation complexity is controlled by selecting \ac{KVI} instances whose definitions require different formula depths and different numbers of \acp{KPI} per \ac{KVI}; the \ac{KVI} scope is controlled by varying the number of requested \ac{KVI} categories; and taxonomy quality \(q\in[0,1]\) is controlled by applying a reproducible degradation procedure to the taxonomy artifacts (e.g., removing a fraction \(1-q\) of entries or fields such as aliases, descriptions, and IDs), thereby modulating how reliable taxonomy grounding is. For each x-axis point, we repeat each method 10 times (different seeds and prompt paraphrases); the plotted curves show the mean and the shaded regions show the variance across runs.

\begin{figure}[t]
    \centering
    \includegraphics[width=0.95\columnwidth]{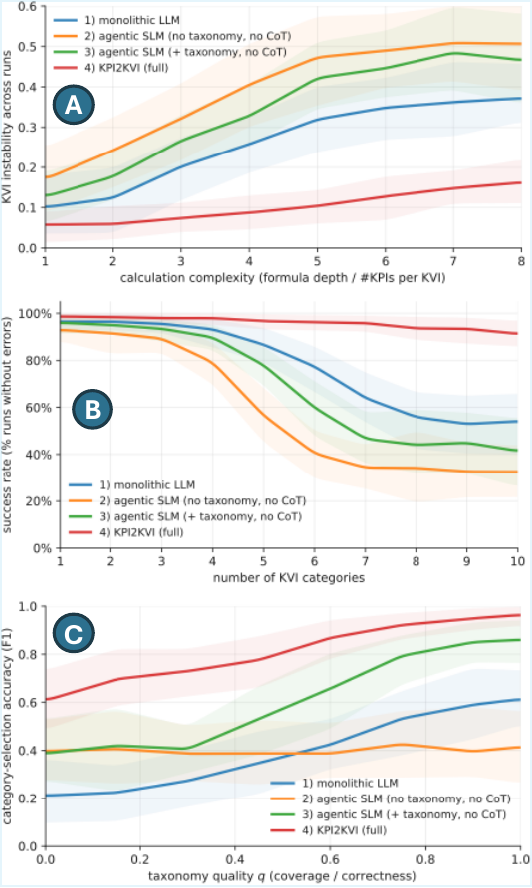}
    \caption{Simulation results for the four method variants. (A) KVI instability across repeated runs versus calculation complexity (formula depth / \#KPIs per KVI). (B) end-to-end success rate (runs without verifier-flagged errors) versus the number of requested KVI categories. (C) category-selection accuracy (F1) versus taxonomy quality \(q\) (coverage/correctness). Curves show mean over 10 runs; shaded regions show variance.}
    \vspace{-0.5cm}
    \label{fig:evaluation}
\end{figure}

Fig.~\ref{fig:evaluation} highlights complementary failure modes. In Fig.~\ref{fig:evaluation}--A, increasing complexity amplifies run-to-run variation in computed \acp{KVI}. KPI2KVI (4) remains the most stable because it structures the \ac{KPI} table, applies explicit calculation steps with bounded outputs, and enables the verifier to catch inconsistencies. The \ac{SLM} baselines without \ac{CoT} (2--3) show higher instability as multi-step arithmetic and unit handling become harder, while the monolithic LLM (1) is typically more stable than the \ac{SLM} variants due to higher model capacity but less stable than the fully staged pipeline because it lacks intermediate structured artifacts and per-\ac{KVI} contracts. In Fig.~\ref{fig:evaluation}--B, the success rate drops as more \ac{KVI} categories are requested, reflecting the difficulty of maintaining a consistent category scope, collecting sufficient \ac{KPI} evidence, and completing all computations at scale. Taxonomy access delays the collapse for the agentic \ac{SLM} (3) relative to (2), and the monolithic LLM (1) degrades more gracefully than the \ac{SLM} baselines but still fails earlier than KPI2KVI (4), which stays robust due to its staged contract (finalized categories), shared memory artifacts, and per-\ac{KVI} computation loop. Finally, Fig.~\ref{fig:evaluation}--C isolates taxonomy dependence: method (2) is largely insensitive to \(q\) (no taxonomy), method (1) improves with \(q\) because the taxonomy is embedded in its prompt, and method (3) benefits strongly from a high-quality taxonomy but can be misled when \(q\) is low; KPI2KVI (4) consistently achieves the best category-selection accuracy by combining taxonomy grounding with the inspector/evaluator refinement loop and explicit finalization of categories.

\section{Conclusion}
In this paper, we presented KPI2KVI, a multi-agent workflow that calculated \acp{KVI} for a given service from its description by translating stakeholder intent into measurable \acp{KPI} and then producing traceable, bounded \ac{KVI} results. We defined a deterministic nine-stage pipeline, specified the responsibilities of each \ac{LLM}-based agent, and implemented an orchestrated architecture with shared memory that stored reusable artifacts such as the interview transcript, the finalized \ac{KVI} category contract, and a structured \ac{KPI} value table. Using taxonomy-grounded category selection and a human-in-the-loop refinement loop, the workflow fixed the \ac{KVI} scope before evidence collection, which improved auditability and reduced ambiguity in downstream calculations. We then computed each \ac{KVI} with explicit formulas, interval bounds, and short rationales that linked outputs to the exact \acp{KPI} and assumptions used. In simulations, KPI2KVI achieved the best overall robustness across increasing calculation complexity, larger requested \ac{KVI} scopes, and varying taxonomy quality. As future work, we planned to fine-tune \acp{SLM} so they no longer required a taxonomy and explicit \ac{CoT} prompting for reliable \ac{KVI} derivation, and to integrate \ac{RAG} to improve long-horizon memory management and reuse of past artifacts across sessions and services.

\section*{Acknowledgment}
The work in this paper was supported in part by the Federal Ministry of Research, Technology, and Space (BMFTR), Germany, through the Project 6GEM+ under Grant 16KIS2411; and in part by the European Union through the 6G-SANDBOX project (Grant No.101096328) and the 6G-Path project (Grant No. 101139172).

\bibliographystyle{IEEEtran}
\bibliography{main}

\end{document}